# Regularities of intermittent luminescence from spherical and tetrapod-shaped quantum dots

A.G. Vitukhnovsky[1], M.M. Kovalev[2], V.V. Lidsky[1*], E.M. Khokhlov[3,4]

[1] *Lebedev Physical Institute, Russian Academy of Sciences*
*Leninskii prospect 53, Moscow, 119991 Russia*

[2] *Moscow State Engineering Physics Institute*
*Kashirskoe sh. 31, Moscow, 115409 Russia*

[3] *Prokhorov General Physics Institute, Russian Academy of Sciences*
*ul.Vavilova 38, Moscow, 119991 Russia*

[4] *Moscow Institute of Physics and Technology*
*Institutskii per. 9, Dolgoprudny, Moscow region, 141700 Russia*

Intermittent photoluminescence of colloidal core/shell semiconductor nanocrystals of spherical and branched shape was studied under CW-laser excitation. Luminescent multichannel registration system was applied for the fluorescence detection of single quantum dots (QDs) in the polystyrene matrix. Comparative statistical data were obtained and analyzed for nano-sphere CdSe/CdS and nano-tetrapod CdTe/CdSe crystals. It was found that "on-" and "off-" blinking times are distributed according to the power law both for spherical CdSe/CdS and tetrapod shape CdTe/CdSe samples in spite of significant QD formfactor differences. We have also found that regardless of the QD-shape both successive "on"-times and successive "off"-times are correlated and pointing out to the memory effect in the mechanism of QD re-emission comprising the prehistory of exciton birth and recombination. Pearson's coefficients were calculated for the correlations of QD-luminescence behavior. Results of experimental measurements were compared with the peculiarities of relevant phenomenon models proposed to date. The data obtained can be used for development of nanophotonic applications and nanocomposite light-emitting materials.

## 1. INTRODUCTION

Colloidal semiconductor nanocrystals (quantum dots 'QD') photophysics has been intensively investigated over the last two decades [1-6]. The reason being that exceptional optical properties of QD, namely broad absorption and narrow emission spectrum, high quantum yield (QY), chemical stability give rise to a vast number of perspective optoelectronic and biological applications: lasers, light-emitting diodes (LEDs) [7], solar batteries [8, 9], biological labels [10, 11] and other promising technologies. In 1996 year Nirmal et al. reveals blinking behavior of single QD by intensity trajectory detection in QD optical microscopy [12]. QD was found undergoes transitions between luminescent state ("on"- state) and nonradiating state ("off"-state) [12-14]. This QD intensity intermittency effect is a great obstacle on the way of the most expected emission applications in the optoelectronics, technology of nanocomposite light-emitting materials, biomedicine [15, 16]. Despite of the numerous works on the matter, the underlying mechanism of such behavior is not understood yet and only semi-empirical models of QD blinking have been proposed to date ([17] and references therein). In particular, the influence of the geometric shape of the nanocrystals on the statistics of intermittent fluorescence is of great interest, since the quantum dot shape modifies electronic structure, carrier dynamics and therefore essential photophysical properties of the nanoparticle [18,19].

---

[*] Electronic address: vlidsky@mail.ru

Newest chemical technologies allow synthesis of colloidal semiconductor nanocrystals of various forms and morphology. Nanospheres, nanorods, nanowires as well as more complicated branched, superbranched and unique colloidal hybrid heterostructures were synthesized in recent years [20-23]. The tetrapod-shaped nanocrystal CdTe/CdSe is the branched nanostructure, in which four rod shaped arms with wurtzite crystal structure branch from a central zinc blende core. CdTe/CdSe QDs of the tetrapod configuration have been investigated in our previous works [24, 25]. Shell-thickness dependences were obtained in the absorption and photoluminescence spectra and decay profiles were measured by time-correlated single photon counting technique. Weak coupling model of quantum wells for branched QD was suggested to explain the kinetics features compared with spherical QDs. Evolution of absorption and photoluminescence spectra of hybrid CdTe/CdSe tetrapod-shaped nanocrystals was studied during CdSe tip growth [23].

In the present work we concentrate on the statistics of fluorescence light re-emitted from single nanocrystals of spherical and tetrapod forms. We report on the first observations of the fluorescence emission from the single tetrpod-shaped nanocrytal CdTe/CdSe, previously studied in our observations of the many-body ensemble of nanoparticles. Power-low distribution for "bright" (or "dark") emission intervals and clear correlation between both successive "on"-time and successive "off"-time periods have been found for these objects.

## 2. EXPERIMENTAL

Colloidal nanocrystal heterostructures of tetrapod and spherical forms have been synthesized by original chemical route at the Faculty of Materials Science of Moscow State University [16]. Typical length and diameter of the CdTe tetrapod arms estimated by TEM were 8 nm and 3 nm, thickness of the CdSe shell was about 1 nm. Diameter of the CdSe spherical core and thickness of the CdS shell were 3 nm and 0.5 nm, respectively. Core/shell QDs CdSe/CdS (spherical form) and CdTe/CdSe (tetrapod form) are shown schematically in Fig.1 along with the electron (hole) density as a function of the nanoparticle radius or the tetrapod arm transverse radial size.

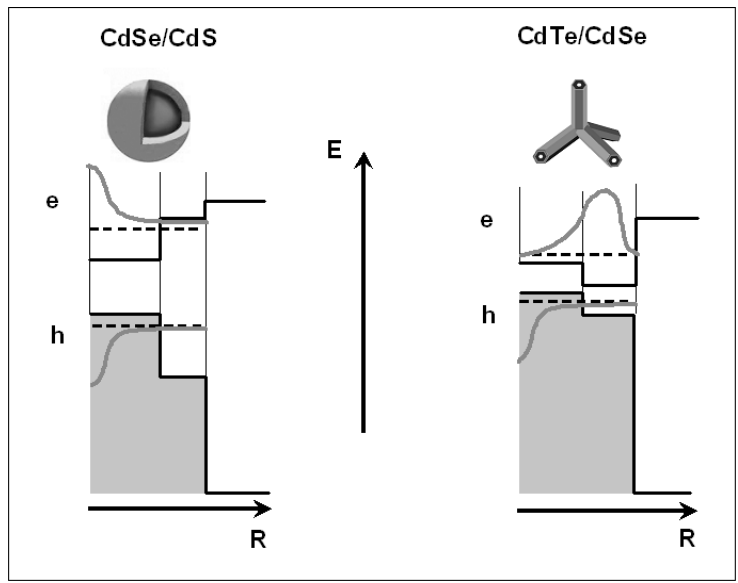

Fig. 1. CdSe/CdS (sphere) and CdTe/CdSe (tetrapod) nanocrystals with type I and type II band alignment. Grey smooth lines depict an electron and hole probability density $|\Psi_{e,h}(r)|^2$ of e,h-localization in the heterostructure nanocrystals.

At the first one (type I) CdS shell of CdSe/CdS acts as an energy barrier for electron (hole), partially isolating it from the surface (Fig.1, left-hand part). Giving the thick enough CdSe shell, energy zone structure of CdTe/CdSe provides predominant location of an electron in the shell and hole in the QD core (type II). This spatial charge separation in CdTe/CdS QDs increases a lifetime of an excited e-h pair, suppresses exciton absorption and increases photoluminescence QY in comparison with ordinary core/shell structures (Fig.1, right-hand part).

The sample preparation procedure briefly consisted of the following stages. Polystyrene diluted in toluene was mixed with QD toluene solution and then spin-coated for 30 sec-onds at 1500 rpm on a glass cover slip, which have been preliminarily thoroughly cleaned chemically to remove trace fluorescent impurities. At the end of drying polymer stage the procedure resulted in 3-5 μm solid film on the glass surface. The particle concentration in the sample was about $10^{10} \div 10^{11}$ см$^{-3}$.

Overview of the experimental apparatus is shown in Fig.2. The excitation radiation of diode laser (L) is expanded by diverging lens (2) to achieve diffraction limited focusing conditions of microscope objective (5). Excitation light attenuated by neutral filter (3) is reflected by dichroic mirror (4) and directed to the sample through objective unit (5). The dichroic mirror both couples the diode laser into the objective as well as separates sample fluorescence and excitation light. Fluorescence from the sample (6) is collected through the microscope objective, filtered with glass combination (7) to remove the illumination of excitation laser (L) and imaged by convergence lens (8) onto receiving aperture of CCD-camera (9). Inset shows the pixel structure of an active detector area, that provides image processing of isolated quantum dot in conjunction with the background signal of its spatial environment. The cooling system was applied to increase the ratio signal/noise CCD by lowering the receiving area temperature to -30°C. Usually about 30 points were observed simultaneously in the field aperture of the receiver. Slow coordinate drift of the selected QDs was compensated by software procedure, the automatic tracking system providing observation of luminescent trail with long-term record up to 150 minutes.

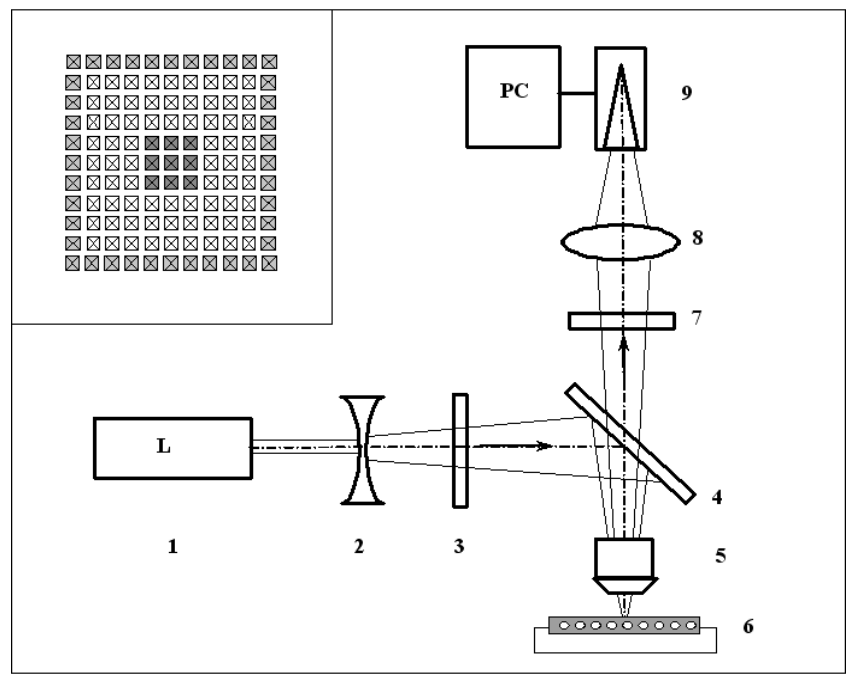

Fig. 2. Experimental setup

1 –CW diode laser ( $\lambda$ = 473 nm), 2 – diverging lens, 3 – neutral filter, 4 – dichroic mirror , 5 – objective, 6 – sample, 7 – notch/colored filter, 8 – convergence lens, 9 – CCD camera, PC computer. The inset presents special pixel arrays to perform the single QD signal processing: 3x3 pixel zone around the center of the QD under observation; intermediate surrounding pixels and peripheral one-pixel contour for the evaluation of background emission.

Temporal behavior of QD radiation intensity $I(t)$ was determined by computational procedure including "signal + background" <S+N> and "background" <N> parts:

$$\langle S+N \rangle = \frac{\sum_i^{k^2} C_i}{k^2}, \qquad (1)$$

where $C_i$ is T-integrated charge from central zone "$i$"-pixel and CCD frame duration T is $25 \cdot 10^{-3}$s;

$$\langle N \rangle = \frac{\left( \sum_i^{n^2} A_i - \sum_j^{m^2} B_j \right)}{n^2 - m^2}, \qquad (2)$$

where $A_i$ is T-integrated "$i$"-pixel charge signal from $n \times n$ total pixel set and $B_j$ is similar signal from $m \times m$ pixels from central and spacer zone. The numerical values of n = 11, m = 9, k = 3 were selected to match optimally optical system configuration. Averaging of difference between components (1) and (2) in the range of four successive frames was a completing operation in forming of fluorescent signal from each individual QD:

$$I(t) \propto \frac{1}{4T} \cdot \sum_i^4 \left( \langle S+N \rangle_i - \langle N \rangle_i \right) \qquad (3)$$

Thus, in our experiments the time dependence of single QD luminescence I(t) was represented in arbitrary units by a series of discrete values defined on above described algorithm with a sampling frequency of 10 Hz. The time series of values accumulated during the hour represents QD intensity trail similar to the one at Fig.3, which illustrated 'blinking' be-havior for both CdSe/CdS and CdTe/CdSe quantum dots under continue laser irradiation. To separate "on"- and "off"- intervals the threshold intensity was taken at the minimum of intensity count histogram for each QD completed trail as shown in Fig. 3.

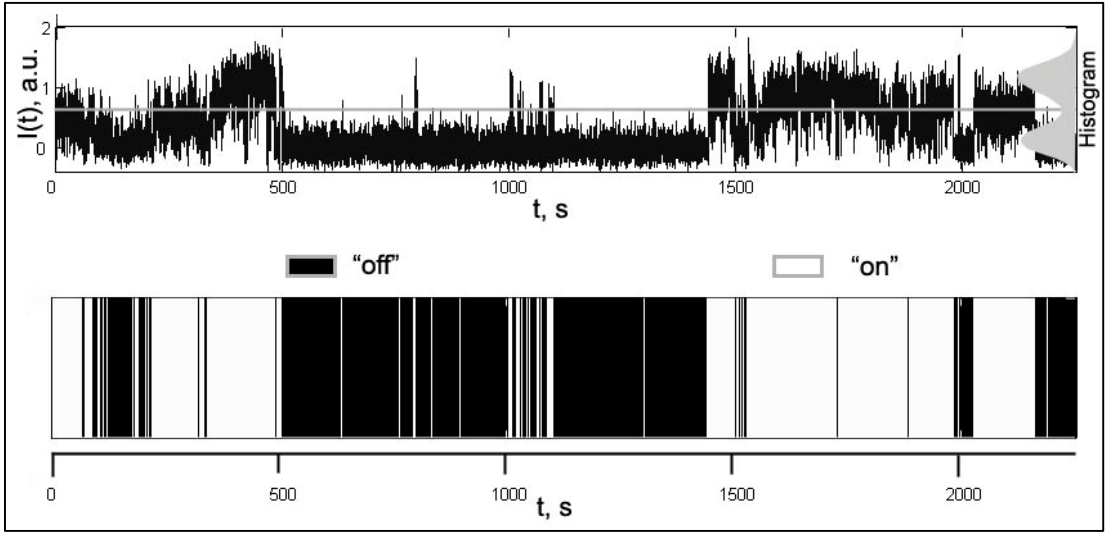

Fig. 3. a) Intensity trail of a single CdSe/CdS QD at the input of pulse-height discriminator and trail-threshold level of photocount histogram; b) fluorescence trail represented by black ("off") and white ("on") time fields resulted from on/off-separation.

The set of $t_{on}$ and $t_{off}$ values derived from a single trail describes the fluorescence statistics of single QDs.

## 3. RESULTS AND DISCUSSION

It is known [26], that single molecules in solid matrix demonstrate similar 'blinking' intensity trails, distributions "on"- and "off"-time intervals being exponential for those luminophores. The conventional theory of single molecule spectroscopy treats phenomenon of bright and dark emission intervals as the time that irradiated molecule spent in the singlet and triplet excited states respectively. It is not the case for single QD [27].

Mechanism of luminescence quenching QD (switch to "off"-state) by the QD-ionization was proposed by discoverers of blinking phenomenon [12]. The followers of this idea have developed detailed model in which the emitting QD is switched "off" when it loses a charge to an external trap-state and extra charge remained quenches fluorescence through rapid nonradiative Auger recombination [13]. Dark QD blinks back "on" when the charge returns from the trap and neutralizes the charged dot. The main feature of such model is the durations of "on" and "off" periods would follow exponential probability densities. The latter, apparently, is caused by three-level kinetic model with one state-trap process [26, 27]. In the meantime nonexponential statistics of "on"/"off "- times could be clearly seen on semi-logarithmic plots of probability density P(t), which represents our measurements of the time-distribution for both type QDs (Fig.4) .

Statistical regularities of QD-fluorescence blinking are presented below in terms of probability density $P(t_{on})$ and $P(t_{off})$ of time-interval obtained from the completed intensity trail of single QD at the context relations:

$$P(t_{on}) \cdot dt = \text{Prob}\{t_{on} \in [t; t + dt]\} \qquad (4)$$

$$P(t_{off}) \cdot dt = \text{Prob}\{t_{off} \in [t; t + dt]\} \qquad (5)$$

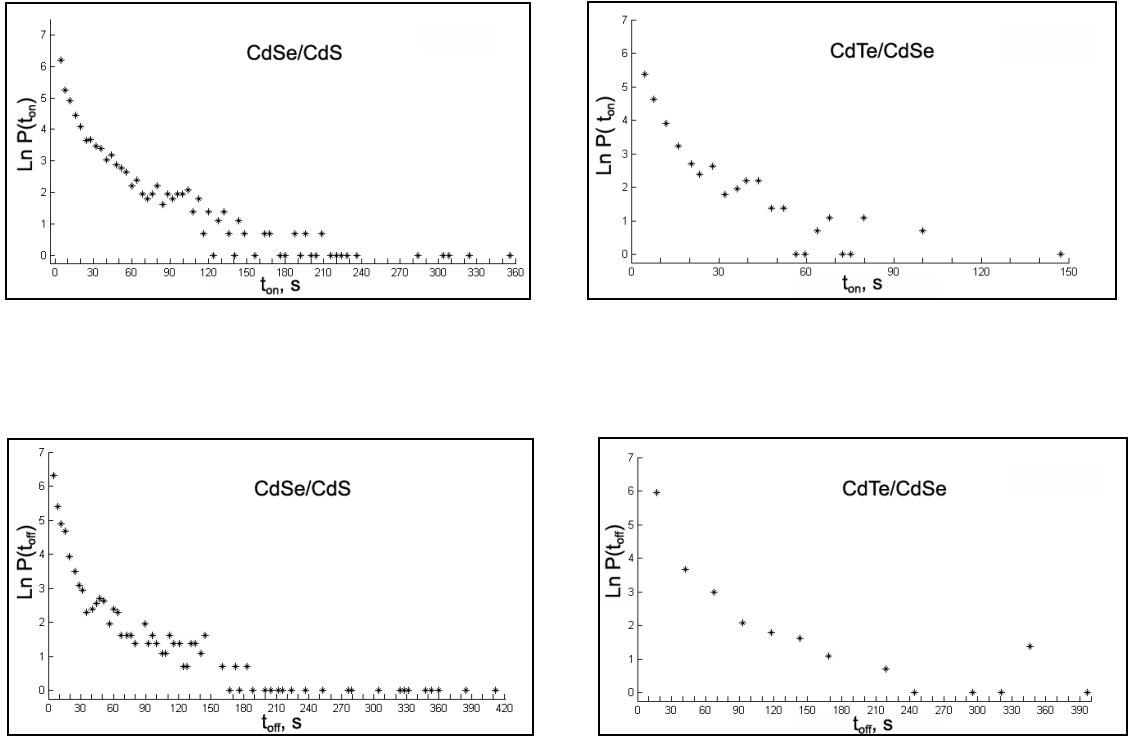

Fig. 4. The top row: natural logarithm of probability density of detecting "on"-interval versus "on" interval length for core/shall CdSe/CdS spherical QDs and CdTe/CdSe tetrapod-shaped QDs. The bottom row: similar dependences for the same samples tested for time "off" - interval distribution.

The plots show that the linearity of ln [P (t)], to be inherent to an exponential distribution, does not hold for both types of time intervals as for spherical dots and for tetrapods. The latter circumstance is very important with respect to relevancy of model assumptions. The ex-planation in the framework of kinetic model of one level-trap is obviously not applicable for QDs in question and necessarily suggests ( at least for "dark" periods) number pathways for recovery from distribution of nonfluorescent QD states [14]. Some sophisticated experiences of such an improvement is presented in [27].

On the other hand, the power law resembles "on"- and "off"- time distributions properly as can be seen in Fig.5. We emphasize the fact of functional similarity P(t) both for CdSe/CdS and CdTe/CdSe QDs:

$$P(t) \propto \frac{1}{t^{-(1+m)}}, \qquad (6).$$

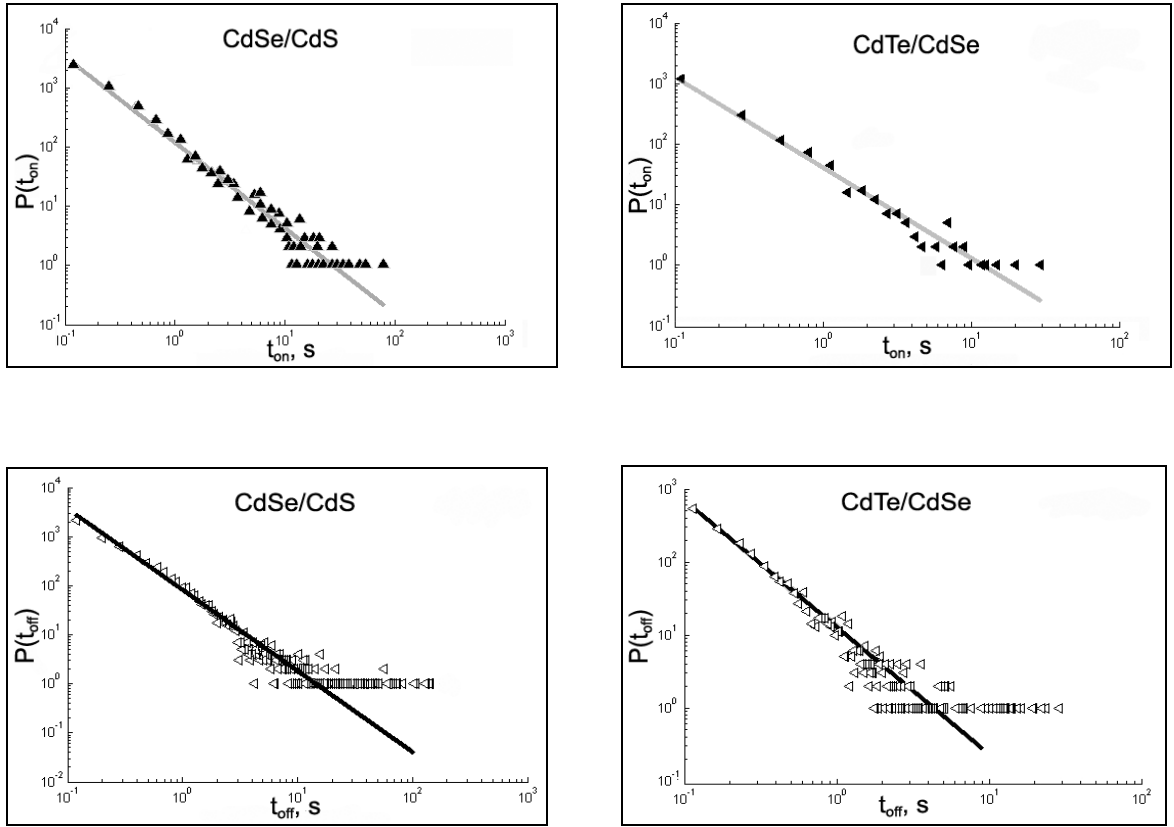

Fig. 5. The top row: natural logarithm of probability density of detecting "on"-interval versus "on"-interval length for core/shall CdSe/CdS spherical QDs (m = 1.49) and CdTe/CdSe tetrapod QDs (m = 1.56). The bottom row: similar dependences for the same samples tested for time "off" interval distribution; (m = 1.72 and m = 1.76, respectively).

We interpret data obtained within the model [28], which takes into account the peculiarities of our observations. This model describes nonexponential distribution of "on" and "off" periods by means of two independent switching processes. The first one gives rise to the both regularities, namely power-low distributed "on" as well as "off" time intervals. The second auxiliary process is exponentially distributed single-rate transition from "on"- to "off"-state. This additional component is responsible for the deviations of $P(t_{on})$ from the power-low ( 6 ), requiring more careful experimental studies.

In addition to these statistical results we investigated correlations of the sequence adjacent events "on"/"on", "off"/"off" and "on"/"off" in the QD fluorescence trail. These characteristics are of great value since they provides information about the microscopic dynamics of quantum dot to test the relevant physical model of the blinking phenomenon. To carry out the analysis of the statistical sets extracted from our experimental date, we used the visual correlations proposed in [12]. The quantitative statistical analysis of this proposal was successively executed in the work of F.D. Stefani et al. [28]. Applying the numerical estimates authors have observed for the first time effect of memory in emission of the spherical nanocrystals ZnCdSe under CW laser irradiation. Detailed and statistically correct distribution of QD-intermittency both "on" and "off" periods in a single model is a solid argument in favor of the theory [28]. Furthermore, recent theoretical findings allow presumably to formulate a reasonable physical mechanism for 'blinking'-memory based on interaction of the multiple recombination centers [29]. Thus, we chose QDs which we could distinguish"on"-and "off" -intervals most accurately for, and constructed scatter plots for the test intervals. Fig. 6 and Fig. 7 depict "on"/"on" and "off"/"off" scatter plots for CdSe/CdS and CdTe/CdSe QDs.

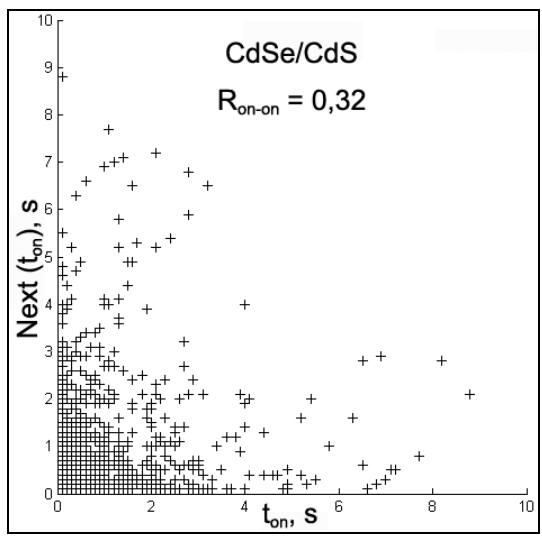
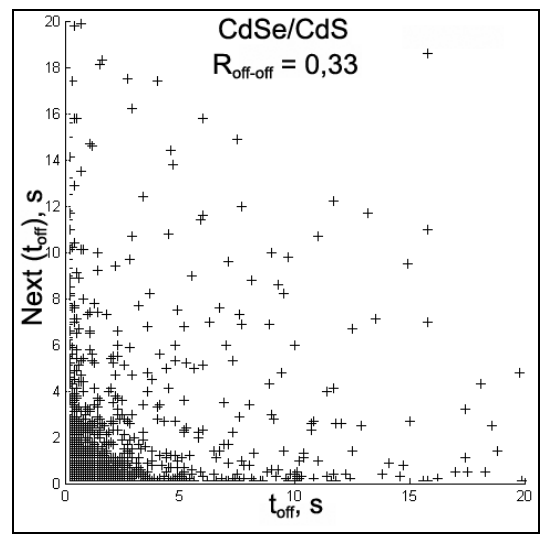

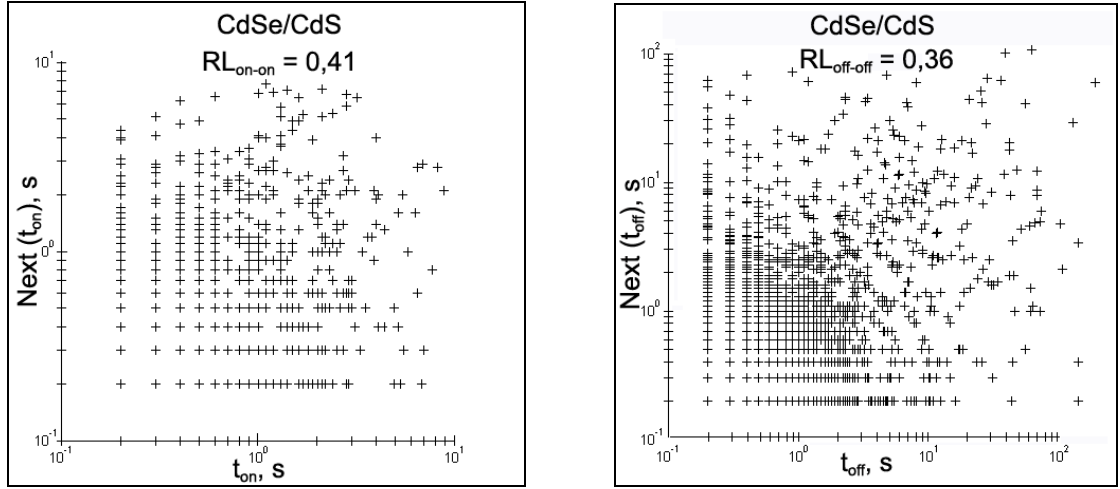

Fig. 6. Scatter plots of successive "on"/ "on" and "off"/ "off" intervals of spherical CdSe/CdS QDs presented in a linear and logarithmic scale.

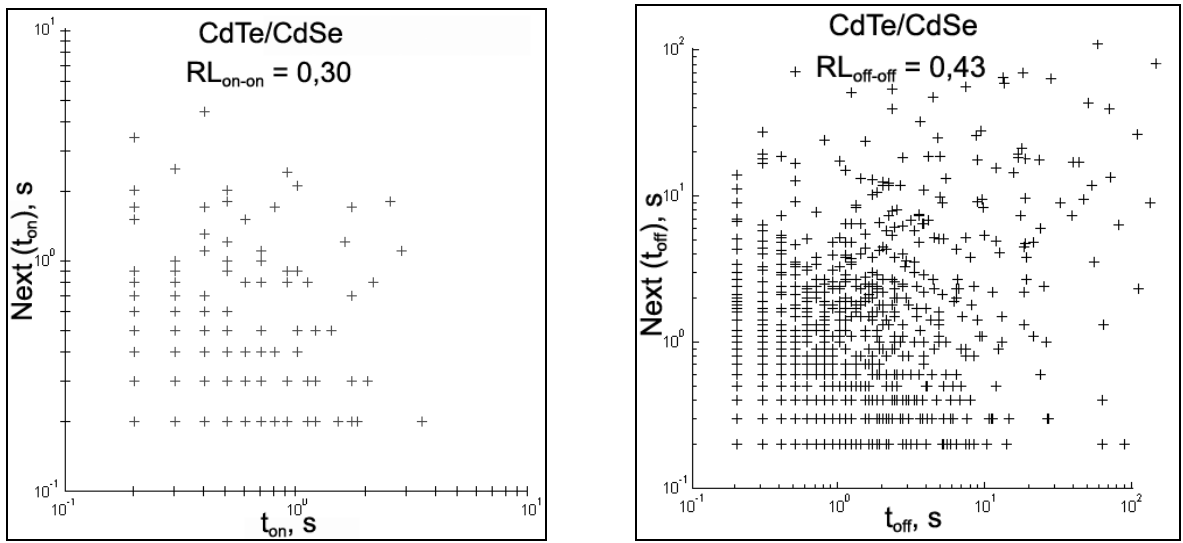

Fig. 7. Scatter plots of successive "on"/"on" and "off"/"off" intervals of CdTe/CdSe tetrapod QDs presented in logarithmic scale. The linear correlation coefficients for the same plots are 0.24 and 0.30 respectively.

Correlation between abscissas and corresponding ordinates is not clearly detectible from the look of the top row of graphs Fig. 6. However, the same scatter plots on a logarithmic scale (lower plots of Fig. 6) reveal the correlation between successive "on"-intervals and between successive "off"-intervals. This feature can be explained by the fact that, as was already shown, $t_{on}$ and $t_{off}$ are power law distributed. Note also that the short $t_{on}$ and $t_{off}$ have axial collineation on logarithmic scale scatter plots due to time line discretization.

In order to describe observed correlations quantitatively, we calculated the linear correlation coefficients R and RL according to Pearson's relations [30]:

$$R \equiv R(X,Y) = \frac{\frac{1}{N} \cdot \sum_{i=1}^{N}(x_i - \bar{x}) \cdot (y_i - \bar{y})}{S_X \cdot S_Y}, \quad (7)$$

where $S_X$ and $S_Y$ are standard deviations:

$$S_X = \sqrt{\frac{1}{N} \cdot \sum_{i=1}^{N}(x_i - \bar{x})^2}, \quad S_Y = \sqrt{\frac{1}{N} \cdot \sum_{i=1}^{N}(y_i - \bar{y})^2}, \quad (8) \text{ and}$$

$$RL \equiv R[(\ln X),(\ln Y)] \quad (9).$$

Here $X_n$ is the length of an interval of prior form ("on" or "off"), and $Y_n$ is the length of adjacent interval of the subsequent form ("on" or "off"), $\bar{x}$ and $\bar{y}$ are their averages. Summation is carried out over all points of each scatter plot. RL in (9) is calculated similarly, but with length of interval being substituted for logarithm of the interval length reduced to dimensionless value by the time unit $\Delta t = 1s$. Possible value of Pearson's coefficient is known to lie on the segment [-1, 1], the values (+1) and (-1) corresponding absolute correlation and anti-correlation between X,Y sets. Zero means absence of a linear correlation between X,Y. The structure of expression RL (9) owing to a logarithmic nonlinearity weighs more the correlation between short time compared with long intervals. In order to be sure that detected correlations are not some features of the experiment or date processing, we also calculated R and RL for a test random signal. Accumulated test and QD statistical data presented in Table1.

| Sample | Pearson's coefficient | Adjacent intervals | |
|---|---|---|---|
| | | "on"/ "on" | "off"/ "off" |
| Random signal | R | -0.0058 | -0.0072 |
| | RL | -0.0069 | -0.0069 |
| CdSe/CdS Sphere | R | 0.32 | 0.33 |
| | RL | 0.41 | 0.36 |
| CdTe/CdSe Tetrapod | R | 0.24 | 0.30 |
| | RL | 0.30 | 0.43 |

Table 1. Correlation coefficients R and RL "on"/ "on" and "off"/ "off" modifications, calculated for a test random signal, CdSe/CdS (sphere) and CdTe/CdSe (tetrapod) quantum dots.

As it is possible to see from the resulted data, R and RL show the considerable correlation between consecutive "on"-intervals and between consecutive "off" - intervals for QDs in question. Analysis of Table 1 shows qualitative agreement of corresponding correlation coefficients for both types of QDs. Taking into account the sharp difference in shape of the objects (sphere and tetrapode), we conclude that the internal mechanism of correlations is an inherent feature of QD regardless of its shape. It should be noted that the observed correlations can not be systematic faults due to the sampling procedure [28], and the apparatus limitation of the signal sampling frequency $F_s$ does not distort the statistics of the interval distribution. We also calculated R and RL for "on"/ "off" adjacent events. R turned out to be -0.06 for CdTe/CdSe QDs and -0.01 for CdSe/CdS QDs, indicating that there is no correlation between "on"-intervals and "off"-intervals.

Pearson's coefficient as a discriminatory element is a "good indicator" of correlated behavior of two random (X,Y) variables while the module of its numerical value is appreciably different from zero. However, the zero value of the Pearson's coefficient can hide a sufficiently rich set of well-structured (X,Y)- patterns described by the nonlinear dependence of random variables. This fact, obviously, should be borne in mind when developing an adequate model, describing the apparent lack of cross-correlations of "on-off" intervals. According to our data obtained there was no noticeable structural organization in the correlation diagrams with zero Pearson's coefficient.

## 4. CONCLUSION

In summary, we have extended fluorescent 'blinking' measurements from individual spherical core/shell semiconductor QDs to the branched tetrapod-shaped nanocrystals. Multichannel fluorescence installation was applied to observe simultaneously the luminescence photons from 30 individual nanoparticles for duration of $10^2 \div 10^4$ s with a sampling frequency of 10 Hz. We have demonstrated for the first time that fluorescence of tetrapod-shaped heterostructures CdTe/CdSe follows the power distribution low of "bright" and "dark" interval just like their spherical CdSe/CdS counterparts. We have also found that regardless of the QD-shape both successive "on"-times and successive "off"-times are correlated and pointing out to the residual memory of 'blinking' process. The experimental measurements are compared with a current model representations describing the main statistical features of intermittent fluorescence of the semiconductor colloidal nanocrystals. The data obtained can be used in development of the nanophotonics elemental base and light-emitting composite materials based on quantum dot technology.

## ACKNOWLEDGMENTS

This work was supported by Russian Foundation for Basic Research (projects 09-02-00546-a). The authors are grateful to A.S. Shul'ga for assistance in obtaining experimental data and useful discussions.

14# REFERENCES

1. Murray C.B., Norris D.J., Bawendi M.G. Synthesis and characterization of nearly monodisperse CdE (E = sulfur, selenium, tellurium) semiconductor nanocrystallites. J. Am. Chem. Soc.,1993, v. 115, p. 8706-8715.

2. Norris D.J., Bawendi M.G. Structure in the lowest absorption feature of CdSe quantum dots. J. Chem. Phys., 1995, v. 103, p. 5260-5268.

3. Alivisatos A.P. Perspectives on the physicalc chemistry of semiconductor nanocrystals. J. Phys. Chem.,1996, v. 100, p. 13226-13239.

4. Peng Z.A., Peng X. J. Nearly monodisperse and shape-controlled CdSe nanocrystals via alternative routes: nucleation and growth. Am. Chem. Soc., 2002, v. 124, p. 3343-3353.

5. Liu H.T., Owen J.S., Alivisatos A.P. Mechanistic study of precursor evolution in colloidal group II-VI semiconductor nanocrystal synthesis. J. Am. Chem. Soc., 2007, v.129, p.305-312.

6. Cho J.W., Kim H.S., Kim Y.J. et al. Ligand-tuned shape control, oriented assembly, and electrochemical characterization of colloidal ZnTe nanocrystals. Chem. Mater., 2008, v. 20, p. 5600-5609.

7. Xu J., Xiao M. Lasing action in colloidal CdS/CdSe/CdS quantum wells. Appl. Phys. Lett., 2005, v. 87, p. 173117-3

8. Earp A.A., Smith G.B, Swift P.D. et al. Maximising the light output of a Luminescent Solar Concentrator. Sol. Energy, 2004, v. 76, p. 655-667.

9. Gur I., Fromer N.A., Alivisatos A.P. Controlled assembly of hybrid bulk-heterojunction solar cells by sequential deposition. J. Phys.Chem. B, 2006, v. 110, p. 25543-25546.

10. Bruchez M. Jr., Moronne M., Gin P. et al. Semiconductor Nanocrystals as Fluorescent Biological Labels. Science, 1998, v. 281, no. 5385, p. 2013-2016.

11. Michalet X., Pinaud F.F., Bentolila L.A. et al. Quantum Dots for Live Cells, in Vivo Imaging, and Diagnostics. Science, 2005, v. 307, no. 5709, p. 538-544.

12. Nirmal M., Dabbousi B.O., Bawendi M.G. et al. Fluorescence intermittency in single cadmium selenide nanocrystals. Nature, 1996, v. 383, p. 802-804.

13. Efros A.L., Rosen M. Random Telegraph Signal in the Photoluminescence Intensity of a Single Quantum Dot. Phys. Rev. Lett., 1997, v. 78, p. 1110-1113.

14. Kuno M., Fromm D.P., Hamann H.F. et al. Nonexponential "blinking" kinetics of single CdSe quantum dots: a universal power law behavior. J.Chem.Phys.,2000, v.112, p.3117-3121.

15. Anikeeva P.O., Halpert J.E., Bawendi M.G. et al. Electroluminescence from a mixed red−green−blue colloidal quantum dot monolayer. Nano Lett., 2007, v. 7, p. 2196-2200.

16. Bates M. A New Approach to Fluorescence Microscopy. Science, 2010, v. 330, no. 6009, p. 1334-1335.

17. Efros A.L.. Nanocrystals: Almost always bright. Nature materials, 2008, v. 7, p. 612-613.

18. Halpert J.E., Porter V.J., Zimmer J.P. et al. Synthesis of CdSe/CdTe Nanobarbells. J Am. Chem. Soc., v. 128, p.13320-13321.